\documentclass[pdflatex,sn-mathphys-num,iicol]{sn-jnl}
\usepackage{graphicx}%
\usepackage{multirow}%
\usepackage{amsmath,amssymb,amsfonts}%
\usepackage{amsthm}%
\usepackage{mathrsfs}%
\usepackage[title]{appendix}%
\usepackage{xcolor}%
\usepackage{textcomp}%
\usepackage{manyfoot}%
\usepackage{booktabs}%
\usepackage{algorithm}%
\usepackage{algorithmicx}%
\usepackage{algpseudocode}%
\usepackage{listings}%
\usepackage{array}%
\usepackage{rotating}
\usepackage[mathlines]{lineno}
\theoremstyle{thmstyleone}%
\theoremstyle{thmstyletwo}%
\theoremstyle{thmstylethree}%
\begin{document}
\title[Article Title]{Thermal conductivity of various CFRPs from 100~mK to 20~K}
\author*[1]{\fnm{Valentin} \sur{Sauvage}}\email{valentin.sauvage@universite-paris-saclay.fr}
\affil*[1]{\orgdiv{Cosmology}, \orgname{Université Paris Saclay, CNRS, Institut d'Astrophysique Spatiale}, \orgaddress{\street{Building 121}, \city{Orsay}, \postcode{90400}, \country{France}}}
\abstract{Carbon-fiber-reinforced polymers (CFRPs) are among the most promising structural materials for aerospace applications, particularly for systems operating at cryogenic temperatures, where high strength, low mass, and low thermal conductivity are essential. In this work, the thermal conductivity of a set of CFRP samples: T300 (67\%~0$^\circ$), T700 (62\%~0$^\circ$), HS40 (67\%~0$^\circ$/0$^\circ$/0$^\circ$/90$^\circ$), HS40 (62\%~0$^\circ$/0$^\circ$/0$^\circ$/90$^\circ$), M55J (0$^\circ$), HS40 (67\%~0$^\circ$), HS40 (62\%~0$^\circ$), IMA (0$^\circ$), and IMA (90$^\circ$), was measured over the temperature range 100~mK to 20~K. Thermal conductivities were extracted using a new analysis method developed. The resulting temperature-dependent behavior for each sample is presented and compared with the existing literature.}
\maketitle
\section{Introduction}
\label{introduction}
Carbon-fiber-reinforced polymers (CFRPs) are playing an increasingly central role in the development of advanced space structures, owing to their combination of low mass, high stiffness, and excellent dimensional stability under severe environmental constraints. Their use has progressively expanded from conventional aerospace components to the highly demanding domain of cryogenic space instrumentation, where structural elements must simultaneously ensure mechanical integrity and stringent thermal isolation. Recent overviews of composite materials for space applications underline this trend, emphasizing the versatility and growing maturity of high-performance CFRPs in orbital and deep-space environments (\cite{tserpes_advances_2025}). At the same time, several space missions, notably the \textit{Herschel} Space Observatory (\cite{mcdonald_thermal_2006}), the \textit{Planck} Space Observatory (\cite{planck_collaboration_planck_2011}), the Mid-Infrared Instrument (MIRI) (\cite{shaughnessy_thermal_2007}) on the \textit{James Webb Space Telescope}, have relied on CFRP struts or support structures precisely because of their reduced thermal conductivity at low temperatures.
However, despite this widespread adoption, the availability of experimental data describing the thermal conductivity of CFRPs at very low temperatures remains limited. Existing measurements typically do not extend below 30--40~K and are often restricted to specific layer configurations or fiber developed for individual missions (\cite{tuttle_cryogenic_2017}). Yet, for modern cryogenic instruments operating between a few kelvin and the millikelvin range, accurate knowledge of the thermal behavior of CFRP components is indispensable. Extrapolating conductivity values from higher-temperature measurements is unreliable in this regime: as our results show, an inflection in the conductivity curves occurs below 1~K, marking a change in the dominant phonon-scattering mechanism. For instruments with cooling budgets at the microwatt level, errors arising from such extrapolation can directly impact the feasibility of the thermal architecture. The thermal transport properties depend sensitively on fibre types, volume fractions, and laminate architecture. In this context, the present work aims to provide a characterization of several CFRP materials between 100~mK and 20~K, thereby furnishing the quantitative data required for refined thermal modeling and for the design of structural elements in upcoming low-temperature space missions.
\section{The thermal conductivity}
The thermal conductivity $\kappa$ is a physical quantity characterizing the material's ability to conduct heat.
It should be noted that, for anisotropic materials such as CFRPs, thermal conductivity is formally a second-order tensor quantity. The present experiment determines an effective one-dimensional conductivity along the sample axis, inferred from a steady-state heat-flow measurement based on Fourier's law. The reported values therefore represent homogenized directional properties of the tested laminate configuration, governed by fiber orientation, stacking sequence, and microstructure, and should not be interpreted as intrinsic material constants~\cite{ke_microstructure_2026}. In all measurements presented in this work, the heat flow is assumed to be strictly one-dimensional, directed along the longitudinal axis of the sample between the two thermometers.
Knowing the thermal conductivity $\kappa$ of a material allows for the calculation of the power $\dot{Q}$ involved in generating a thermal gradient $T_2$ to $T_1$ along a length $L$ through a cross-section $S$. This relation is given by equation~\ref{eq:theoretical_thermal_conductivity_equation}.
\begin{equation}
    \dot{Q} = \frac{S}{L}\int_{T_1}^{T_2} \kappa (T)\,dT
    \label{eq:theoretical_thermal_conductivity_equation}
\end{equation}
\begin{tabular}{>{\raggedleft\arraybackslash}p{0cm} >{\raggedright\arraybackslash}p{0cm} >{\raggedright\arraybackslash}p{5cm} >{\raggedleft\arraybackslash}p{2cm}}
	$\dot{Q}$ & : & the power flowing through the sample & [W] \\
    $S$ & : & the sample section & [m$^2$] \\
    $L$ & : & the distance between the two thermometers & [m] \\
    $T_1$ & : & the bottom temperature & [K] \\
    $T_2$ & : & the top temperature & [K] \\
    $\kappa$ & : & the sample thermal conductivity & [W.m$^{-1}$.K$^{-1}$] \\
\end{tabular}

This equation defines the way the thermal conductivity $\kappa$ will be determined using the method presented in section~\ref{sec:method_ias}.
\section{Experimental setup}
\label{subsec:experimental_setup_ias}
\begin{figure}
	\centering
	\includegraphics[width=\columnwidth]{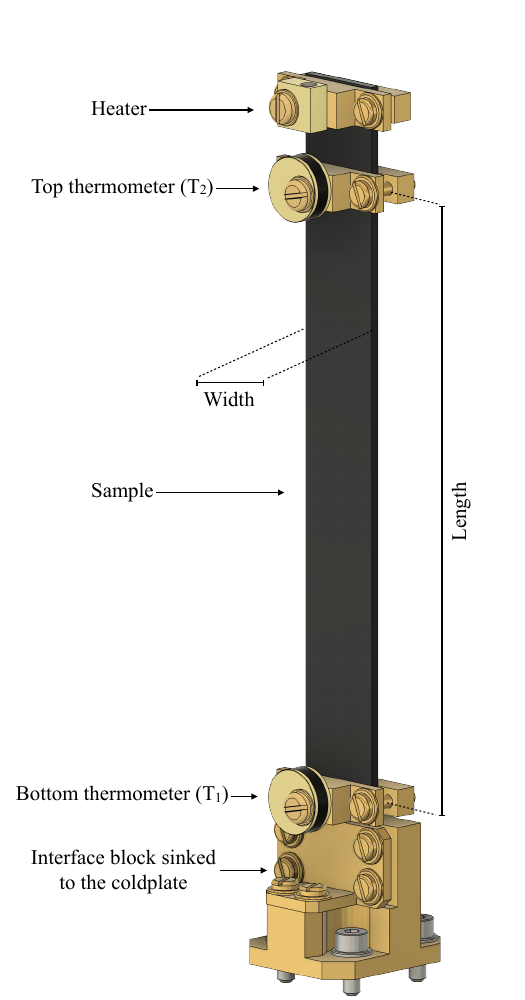}
	\caption{The instrumented sample was placed on the cold plate.}
	\label{fig:experimental_setup}
\end{figure}

We developed an experimental setup designed to extract the thermal conductivity $\kappa$ consistent with the parameters required in equation~\ref{eq:theoretical_thermal_conductivity_equation}. This configuration was implemented at the Institut d'Astrophysique Spatiale within the DRACuLA cryostat described in \cite{sauvage_development_2023}. The system is a BlueFors LD400 dilution refrigerator capable of reaching temperatures below 10~mK, with a cooling power of 500~$\mu$W at 100~mK. The cryostat provides a 290~mm diameter cold plate and an available height of 400~mm, allowing for mounting several samples simultaneously. The radiative environment during the measurements is approximately 1~K.
Figure~\ref{fig:experimental_setup} represents the sample fully instrumented while placed on the cold plate.
\subsection{Thermometry}
\begin{figure}
	\centering
	\includegraphics[width=\columnwidth]{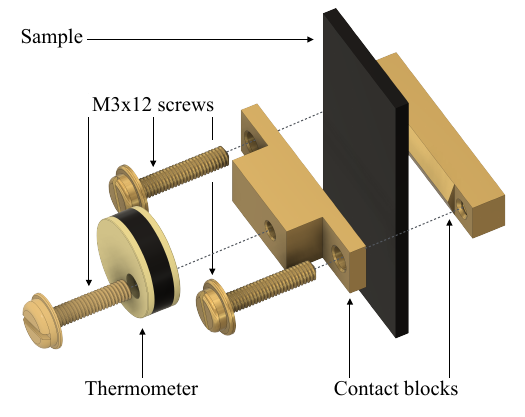}
	\caption{Exploded view of a ``linear'' contact for both top and bottom thermometers. The choice of a linear contact was made to reduce the uncertainties of the length between the two thermometers.}
	\label{fig:sample_contacts_setup}
\end{figure}

\begin{figure}
	\centering
	\includegraphics[width=\columnwidth]{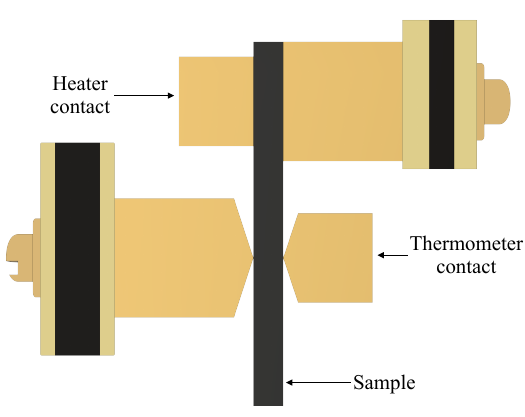}
	\caption{On top the contact for the heater to maximize the injected power into the sample, and on bottom the linear contact for the thermometry to limit the uncertainties of the length between the two thermometers.}
	\label{fig:sample_cross_section}
\end{figure}

Two thermometers are used to determine the thermal conductivity $\kappa$ of the sample, distant by a length $L$. One is on top of the sample and provides $T_2$, and one is at the bottom of the sample and provides $T_1$. The two thermometers need to be in contact with the sample; a dedicated system has been designed to ensure this contact, as shown in Figure~\ref{fig:sample_contacts_setup}. Also, to reduce the uncertainties related to the length between the two thermometers, the contact was chosen to be linear and not surface, as represented in Figure~\ref{fig:sample_cross_section}. The thermometers are mounted by M3$\times$12 brass screws with a tightening torque of 0.61~N.m.
The bottom thermometer is a Cernox RX-102A (range: [0.05~K--100~K]), and the top thermometer is a Cernox CX-1010 (range: [0.1~K--100~K]). Both are calibrated by the author inside the DRACuLA cryostat \cite{sauvage_development_2023} against the ROx main thermometer of the cryostat. The thermometry is monitored by a Lakeshore 370 AC resistance bridge with an excitation voltage that agrees with the manufacturer's recommendations. The uncertainty of the measured $T_1$ and $T_2$ is estimated to be 1\%. This value accounts for two contributions: the calibration uncertainty of the Cernox thermometers, established by in-situ calibration against the cryostat ROx reference thermometer over the full measurement temperature range including sub-Kelvin temperatures, and the noise floor of the Lakeshore 370 AC resistance bridge at the operating excitation level. Calibration drift due to thermal cycling and hysteresis effects were not independently quantified in this campaign; however, manufacturer specifications for Cernox sensors indicate that such effects are typically below 0.5\%, which falls within the quoted 1\% uncertainty budget.
Both thermometers are connected by four phosphor-bronze 127~$\mu$m wires of 400~mm to the cold-plate cryostat connectors bracket.
\subsection{Heater}
\begin{figure}
	\centering
	\includegraphics[width=\columnwidth]{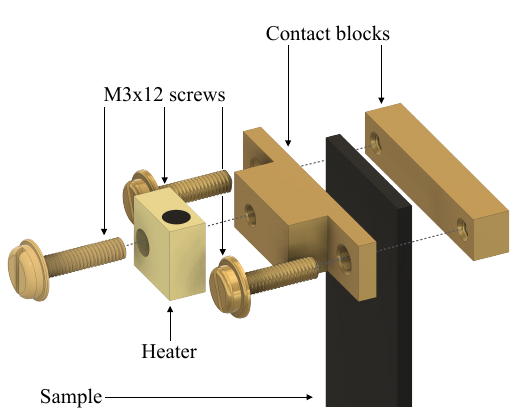}
	\caption{Exploded view of the interface design to thermally couple the heater and the sample.}
	\label{fig:sample_contact_heater}
\end{figure}

To generate the thermal gradient along the length, a heater is placed on top of the sample. The heater is composed of a 10~k$\Omega$ thin metallic film resistor (with low $\frac{\Delta R}{\Delta T}$) confined into a gold-plated copper block with Stycast 2850-FT. The heater is thermally connected to the sample by a dedicated design shown in Figure~\ref{fig:sample_contact_heater}, where all M3$\times$12 brass screws were tightened with a torque of 0.61~N.m.
The heater is connected to the cold-plate cryostat connectors bracket by four phosphor-bronze 127~$\mu$m wires of a 400~mm minimum length. A current is injected through the resistance, and the voltage is monitored at both ends, by a Keithley 2602A. This method provides the true power injected, $\dot{Q}$, through the sample. The uncertainty of the measured $\dot{Q}$ is assumed to be 0.1\% according to the recent instrumental calibration of the Keithley 2602A, which employs a four-wire configuration eliminating lead resistance contributions. At the lowest power levels (below $\sim$10~nW), this estimate represents a lower bound; however, the power uncertainty remains subdominant with respect to the 1\% thermometric uncertainties.
\subsection{Interface with the cryostat}
\begin{figure}
	\centering
	\includegraphics[width=\columnwidth]{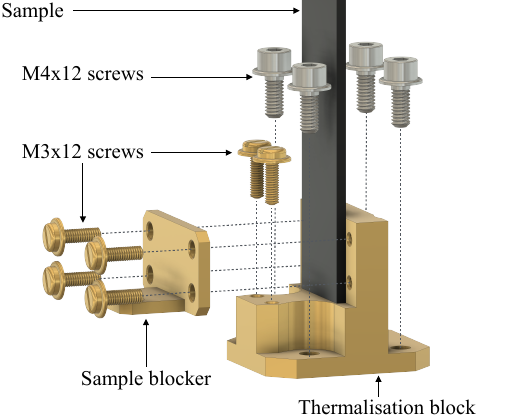}
	\caption{Mechanical design to anchor the sample on the cold plate of the cryostat. To ensure good thermal contact, the interface is a large surface between the sample and the copper gold-plated cold plate.}
	\label{fig:contact_sample_cold_plate}
\end{figure}

The sample needs to be anchored on the cold plate for proper thermalization. To do that, the design presented in Figure~\ref{fig:contact_sample_cold_plate} is used. Four A2-70 M4$\times$12 screws are used to anchor the interface to the cold plate. Both contact surfaces have been cleaned, and a torque of 1.4~N.m was applied. Then, the sample is clamped between the thermalization block and the sample blocker with M3$\times$12 brass screws (torque 0.61~N.m).
\section{Method}
\label{sec:method_ias}
\subsection{Determination of $T_1$, $T_2$ and $\dot{Q}$}
\begin{figure}
	\centering
	\includegraphics[width=\columnwidth]{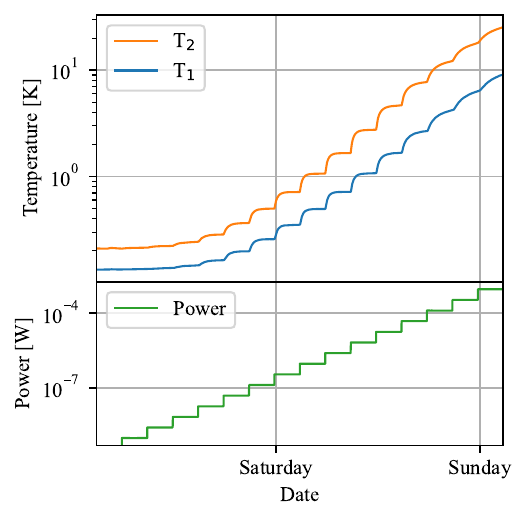}
	\caption{Top plot showing the temperature evolution of the two thermometers and the injected power on bottom plot against the time.}
	\label{fig:plot_temperature_power}
\end{figure}

\begin{figure}
	\centering
	\includegraphics[width=\columnwidth]{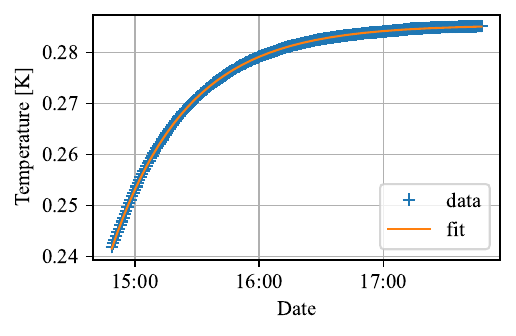}
	\caption{Example of the fit of one step using equation~\ref{eq:temperature_fit_step}. The data are represented together with the fit.}
	\label{fig:plot_fit_temperature}
\end{figure}

\begin{figure}
	\centering
	\includegraphics[width=\columnwidth]{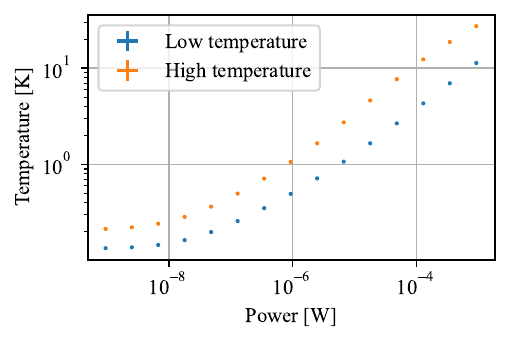}
	\caption{Graphical representation of both temperature measurements against the injected power, together with their uncertainties.}
	\label{fig:results_all_temperatures}
\end{figure}

\begin{figure}
	\centering
	\includegraphics[width=\columnwidth]{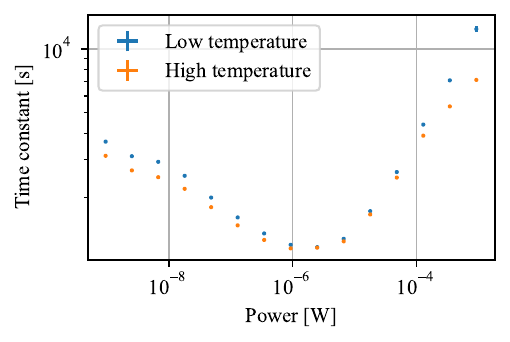}
	\caption{Graphical representation of the time constant for both thermometers against the injected power, together with their uncertainties.}
	\label{fig:results_all_time_constant}
\end{figure}

\begin{figure}
	\centering
	\includegraphics[width=\columnwidth]{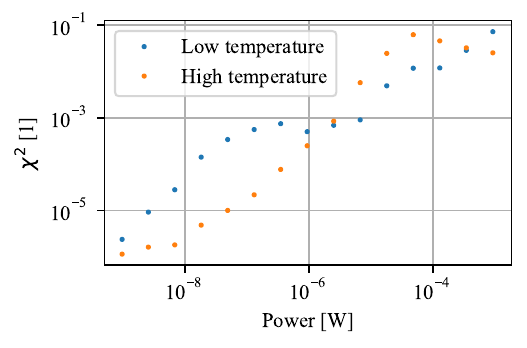}
	\caption{Graphical representation of the $\chi^2$ for every fit performed (lower is better).}
	\label{fig:results_all_chi2}
\end{figure}

In order to determine the thermal conductivity $\kappa$ experimentally, a sequence of various powers from nW to mW (defined as ``step'') is injected at one end of the sample (Figure~\ref{fig:plot_temperature_power}, bottom plot), and the temperature of both thermometers is monitored (Figure~\ref{fig:plot_temperature_power}, top plot). Each step lasted 3 hours.
From the raw data, the temperature and the time constant are extracted. Each step is fitted with equation~\ref{eq:temperature_fit_step}.
\begin{equation}
    T(t) = A - Be^{-\frac{t}{\tau}}
    \label{eq:temperature_fit_step}
\end{equation}
\begin{tabular}{>{\raggedleft\arraybackslash}p{0cm} >{\raggedright\arraybackslash}p{0cm} >{\raggedright\arraybackslash}p{5cm} >{\raggedleft\arraybackslash}p{2cm}}
	$T$ & : & the temperature against time & [K] \\
    $A$ & : & a parameter & [K] \\
    $B$ & : & a parameter & [K] \\
    $t$ & : & the time & [s] \\
    $\tau$ & : & the time constant & [s] \\
\end{tabular}
The parameter $A$, the final temperature reached at equilibrium (as $t\to\infty$), can be obtained from equation~\ref{eq:temperature_fit_step}. The parameter $B$ represents the initial temperature offset such that $T(0) = A - B$. Hence, there is no need to wait for the temperature stabilization of the system. In return, an uncertainty is associated with the final temperature and is taken into account for the rest of this paper. Figure~\ref{fig:plot_fit_temperature} shows an example of a step together with its associated fit. This method provides $T_1$ and $T_2$.
The average of the measured power is designated as $\dot{Q}$, while its standard deviation over the duration of the step is considered as the associated uncertainty. The fitting method is applied for each step, and the results are shown in Figure~\ref{fig:results_all_temperatures}, where $T_1$ and $T_2$ are represented against the injected power, together with their measurement and fit uncertainties.
We can also provide the time constant for every measured step, represented in Figure~\ref{fig:results_all_time_constant}. The time constant value reaches a minimum around 1~$\mu$W, which corresponds to approximately 1~K, which is due to a trade-off between the thermal conductivity and the specific heat. Note that the specific heat will not be determined in this paper.
For each fit, the goodness of fit is assessed using the $\chi^2$ method, with the results shown in Figure~\ref{fig:results_all_chi2}. This evaluation provides a measure of how well the model corresponds to the data. The lower the $\chi^2$ value, the better the fit. Given the number of data points measured for each step, a $\chi^2$ value below 10 is deemed excellent.
\subsection{Determination of $\dot{Q}_0$}
Since equation~\ref{eq:theoretical_thermal_conductivity_equation} is theoretical and does not account for the environmental setup, it is incomplete. To address this, a parameter $\dot{Q}_0$ must be introduced. This parameter represents an additional heating originating from the environment, which generates a constant thermal gradient. The influence of $\dot{Q}_0$ becomes significant when the injected power $\dot{Q}$ is low (below 100~nW) and must be subtracted from the data to accurately determine the true $\dot{Q}$ flowing through the sample. Equation~\ref{eq:experimental_thermal_conductivity_equation} is considered to give the thermal conductivity $\kappa$ of the sample.
\begin{equation}
    \dot{Q} = \frac{S}{L}\int_{T_1}^{T_2} \kappa (T)\,dT + \dot{Q}_0
    \label{eq:experimental_thermal_conductivity_equation}
\end{equation}
The parameter $\dot{Q}_0$ was fitted independently for each sample. Its value was verified to remain consistent across the low-power steps of a given run, confirming that environmental parasitic heating was stable over the multi-day duration of each measurement campaign.
\subsection{Determination of the thermal conductivity}
Then, an algorithm is applied to determine the thermal conductivity $\kappa$, using the \textit{optimize.curve\_fit} function from the \textit{scipy} Python package. The functional form of the thermal conductivity $\kappa(T)$ must be provided. A simple power-law model $\kappa \propto T^\alpha$ cannot simultaneously capture the low-temperature $T^3$ Debye regime and the weaker temperature dependence observed at higher temperatures. The Callaway-type~\cite{callaway_model_1959} form of equation~\ref{eq:fit_function_thermal_conductivity} encodes both regimes in a single expression and is therefore better suited to the full measurement range of 100~mK to 20~K.
\begin{equation}
    \kappa(T) = \frac{a T^3}{1 + bT^n}
    \label{eq:fit_function_thermal_conductivity}
\end{equation}
\begin{tabular}{>{\raggedright\arraybackslash}p{1.2cm}%
                >{\raggedright\arraybackslash}p{0.3cm}%
                >{\raggedright\arraybackslash}p{3.5cm}%
                >{\raggedleft\arraybackslash}p{2cm}}
    $\kappa$     & : & the thermal conductivity  & [W.m$^{-1}$.K$^{-1}$] \\
    $T$          & : & the temperature           & [K] \\
    $a, b, n$    & : & fitting parameters        & [--] \\
\end{tabular}

\begin{figure}
	\centering
	\includegraphics[width=\columnwidth]{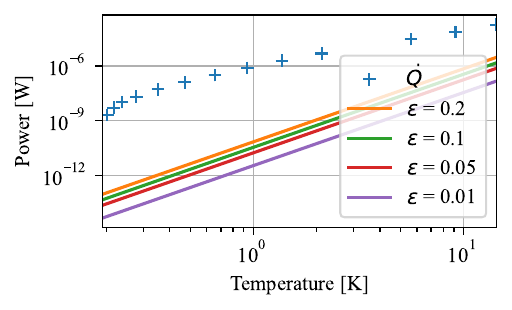}
	\caption{Comparison between the injected power $\dot{Q}$ and the radiative power loss predicted by the model, evaluated for different surface emissivities and temperatures.}
	\label{fig:comparison_power_loss_radiation}
\end{figure}

The $T^3$ numerator reflects the low-temperature Debye phonon density of states in the boundary- and defect-scattering-limited regime, while the denominator term $bT^n$ accounts for the onset of additional scattering channels, such as fibre-matrix interface scattering and the contribution of the epoxy matrix, at higher temperature.
This model does not include the radiative power loss associated with the emissivity of the sample, which scales as $T^{4}$ and becomes relevant whenever the surrounding radiative environment (i.e.\ the shield temperature) is colder than the sample itself. At higher temperatures, the associated emissive loss may become non-negligible compared to the applied heating power, and the model must therefore be extended accordingly. A quantitative estimate, as shown in Figure~\ref{fig:comparison_power_loss_radiation},  indicates that, for emissivities representative of the tested samples ($\epsilon \sim 0.05$--$0.1$), radiative losses exceed 1\% of the injected power only above approximately 15--20~K, which corresponds to the upper boundary of the measurement range. Below this threshold, the radiative contribution remains safely negligible.
Since the injected power covers several orders of magnitude (from $10^{-9}$ to $10^{-3}$~W), the convergence criterion of the \textit{optimize.curve\_fit} routine, namely the least-square tolerance set at $10^{-6}$, is not appropriate for such small fitted values. The fit must therefore be performed in logarithmic space. In order to compute the residuals, the error estimation has also been determined so that the relative weight of the error remains identical, independently of the power, as expressed in equation~\ref{eq:error_log}.
\begin{equation}
\label{eq:error_log}
\text{error} = \sqrt{\frac{1}{N} \sum_{i=1}^{N} \log\!\left(\frac{\dot{Q}^{\text{fit}}_i}{\dot{Q}^{\text{meas}}_i + \dot{Q}_0}\right)^{\!2}}
\end{equation}
\begin{tabular}{>{\raggedright\arraybackslash}p{0.5cm} >{\raggedright\arraybackslash}p{0cm} >{\raggedright\arraybackslash}p{4.5cm} >{\raggedleft\arraybackslash}p{2cm}}
	$\dot{Q}^{\text{fit}}$  & : & the power fitted with $\kappa$ & [W] \\
    $\dot{Q}^{\text{meas}}$ & : & the power injected on sample   & [W] \\
    $\dot{Q}_0$             & : & the parasitic heat             & [W] \\
    $N$                     & : & the number of measurements     & [1] \\
\end{tabular}
\subsection{Source of uncertainties}
This experimental determination method involves several sources of uncertainty, each associated with its respective parameter.
An uncertainty of 0.1\% is associated with $\dot{Q}$, coming from the power measurement. Uncertainties of 1\% are associated with $T_1$ and $T_2$, due to thermometer calibration and electronic read-out.
Uncertainties of 0.1\% are associated with $L$ and $S$, resulting from dimensional measurements (measurements available in table~\ref{tab:cfrp_sample_geometric_dimentions}). The thermal contraction of the sample induces measurable changes in its geometrical dimensions. Because of the preferential fibre orientation, the CFRP exhibits pronounced anisotropy \cite{sapi_properties_2020}, which renders the determination of an effective Coefficient of Thermal Expansion (CTE) for the composite particularly challenging. At cryogenic temperatures, the CTE of carbon fibres is known to be very small \cite{srisuriyachot_quantification_2025}, whereas that of the epoxy matrix binding the fibres may reach 1.2 to 1.3\% for an unplasticized resin system~\cite{nakane_thermal_2002}. Based on this measured contraction of the unplasticized epoxy matrix, a uniform correction of 1.3\% has therefore been applied to all geometrical dimensions. It should be noted that this uniform correction is an approximation. Due to the pronounced CTE anisotropy of CFRPs, the longitudinal contraction (along the fibre direction) is expected to be smaller than the transverse contraction. These effects partially compensate in the ratio $S/L$ entering equation~\ref{eq:theoretical_thermal_conductivity_equation}. In addition, Baschek et al.~\cite{baschek_parameters_1998} reported a variation of approximately 20\% in the CTE of samples subjected to thermal cycling, which is not considered in this study.
The power dissipated through the wires was evaluated and found to be consistently negligible compared with the injected power~$\dot{Q}$. The uncertainty associated with radiative losses has already been addressed previously.
All these uncertainties were propagated using a Monte Carlo approach with 200 iterations. They were fully incorporated into the fitting procedure so as to obtain the most reliable estimate of the thermal conductivity $\kappa$ together with its overall uncertainty. The residuals between the fitted curve and the experimental data define the final error associated with the method (equation~\ref{eq:error_log}). The resulting thermal conductivities are shown in Figure~\ref{fig:all_conductivities}, and the corresponding fitting parameters are reported in table~\ref{tab:fit_parameters}.
\section{Results}
\begin{figure*}[t]
	\includegraphics[width=\textwidth]{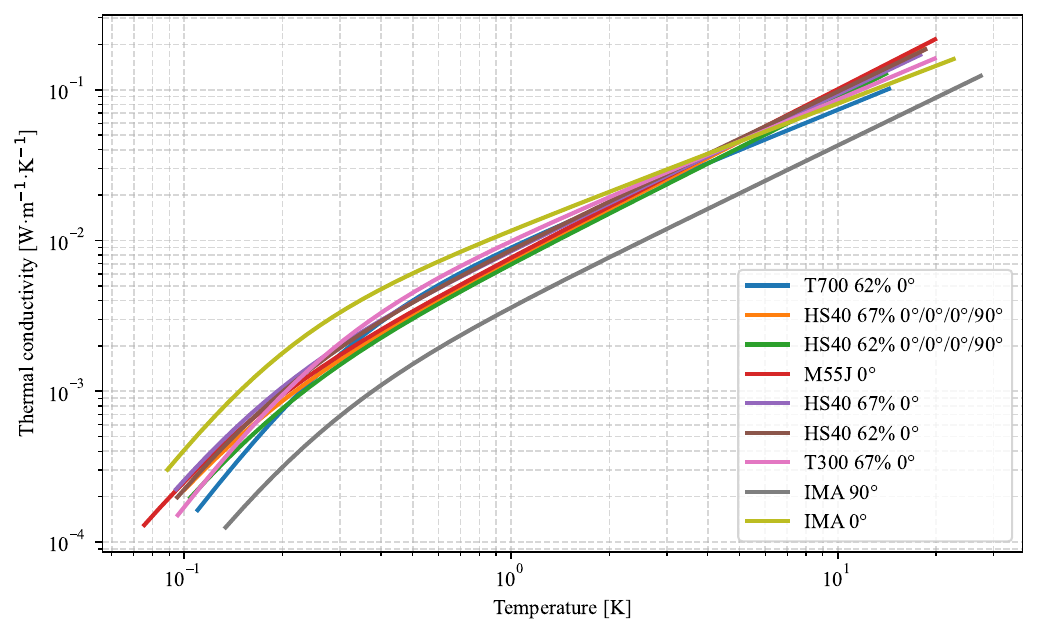}
	\caption{Thermal conductivity $\kappa$ of the nine samples determined at IAS over the temperature range [100~mK -- 20~K].}
	\label{fig:all_conductivities}
\end{figure*}

Owing to their expected low thermal conductivity and specific heat, the samples required more than three days to reach thermal equilibrium at the lowest attainable temperature. The samples were mounted in groups of three on the cold plate, but tested individually in order to avoid any mutual parasitic radiative load.
We tested nine commercially available CFRP samples, differing in fiber type, fiber content (volume ratio between fiber and total volume), and fiber orientation, which are summarized in Table~\ref{tab:cfrp_sample_characteristics}. The fiber orientation refers to the angle between the fiber direction and the heat flux: a fiber aligned with the flux is denoted 0$^\circ$, whereas a transverse fiber is noted 90$^\circ$.
The CFRP samples were provided as elongated rectangular plates approximately 200~mm in length, 20~mm in width, and 2~mm in thickness. Precise measurements of the geometrical dimensions used in the thermal-conductivity analysis, together with their absolute uncertainties, are given in Table~\ref{tab:cfrp_sample_geometric_dimentions}.
\begin{table}[ht]
    \centering
    \renewcommand{\arraystretch}{1}
    \begin{tabular}{|c||c|c|c|c|}
        \hline
        Sample & Fiber & Fiber content & Angle [$^\circ$] & Supplier \\ \hline\hline
        1 & T700 & 62\% & 0          & NTPT      \\ \hline
        2 & HS40 & 67\% & 0/0/0/90   & NTPT      \\ \hline
        3 & HS40 & 62\% & 0/0/0/90   & NTPT      \\ \hline
        4 & M55J & 60\% & 0          & Mecano ID \\ \hline
        5 & HS40 & 67\% & 0          & NTPT      \\ \hline
        6 & HS40 & 62\% & 0          & NTPT      \\ \hline
        7 & T300 & 67\% & 0          & NTPT      \\ \hline
        8 & IMA  & 60\% & 90         & Mecano ID \\ \hline
        9 & IMA  & 60\% & 0          & Mecano ID \\ \hline
    \end{tabular}
    \caption{Details of the CFRP samples tested are presented. The commercial designation, fiber content, fiber-orientation angle, and supplier are provided.}
    \label{tab:cfrp_sample_characteristics}
\end{table}
\begin{table}[ht]
    \centering
    \renewcommand{\arraystretch}{1}
    \begin{tabular}{|c||c|c|c|}
        \hline
        Sample & Length [mm] & Width [mm] & Thickness [mm] \\ \hline\hline
        1 & 144.3 $\pm$ 0.1 & 20.16 $\pm$ 0.01 & 1.97 $\pm$ 0.02 \\ \hline
        2 & 143.5 $\pm$ 0.1 & 20.16 $\pm$ 0.01 & 1.97 $\pm$ 0.02 \\ \hline
        3 & 143.5 $\pm$ 0.1 & 20.16 $\pm$ 0.01 & 1.97 $\pm$ 0.02 \\ \hline
        4 & 55.3  $\pm$ 0.1 & 20.85 $\pm$ 0.01 & 1.97 $\pm$ 0.02 \\ \hline
        5 & 155.5 $\pm$ 0.1 & 20.16 $\pm$ 0.01 & 1.97 $\pm$ 0.02 \\ \hline
        6 & 156.2 $\pm$ 0.1 & 20.16 $\pm$ 0.01 & 1.97 $\pm$ 0.02 \\ \hline
        7 & 156.1 $\pm$ 0.1 & 20.16 $\pm$ 0.01 & 1.97 $\pm$ 0.02 \\ \hline
        8 & 65.8  $\pm$ 0.1 & 20.16 $\pm$ 0.01 & 1.97 $\pm$ 0.02 \\ \hline
        9 & 65.5  $\pm$ 0.1 & 20.16 $\pm$ 0.01 & 1.97 $\pm$ 0.02 \\ \hline
    \end{tabular}
    \caption{CFRP sample geometric dimensions.}
    \label{tab:cfrp_sample_geometric_dimentions}
\end{table}
Applying the method described in section~\ref{sec:method_ias}, we report the fitting parameters of the thermal conductivity $\kappa$ model given in equation~\ref{eq:fit_function_thermal_conductivity}, together with their relative uncertainties and the corresponding temperature ranges. The resulting thermal conductivities are shown in Figure~\ref{fig:all_conductivities}.
\begin{sidewaystable*}
\centering
\caption{Fitted thermal conductivity parameters for all samples. Values are mean $\pm$ standard deviation from Monte Carlo resampling. Errors are relative RMS values expressed in percent.}
\label{tab:fit_parameters}
\begin{tabular}{lccccc}
\hline
Sample & $a$ & $b$ & $n$ & Error [\%] & Temperature range [K] \\
\hline
T700 62\% 0$^\circ$ & 0.140 $\pm$ 0.003 & 14.580 $\pm$ 0.474 & 2.114 $\pm$ 0.016 & 2.44 & [0.110 -- 14.347] \\
HS40 67\% 0$^\circ$/0$^\circ$/0$^\circ$/90$^\circ$ & 0.364 $\pm$ 0.028 & 48.723 $\pm$ 4.591 & 1.880 $\pm$ 0.013 & 4.09 & [0.100 -- 18.206] \\
HS40 62\% 0$^\circ$/0$^\circ$/0$^\circ$/90$^\circ$ & 0.258 $\pm$ 0.039 & 36.157 $\pm$ 6.791 & 1.908 $\pm$ 0.032 & 2.19 & [0.104 -- 14.044] \\
M55J 0$^\circ$ & 0.427 $\pm$ 0.017 & 54.647 $\pm$ 2.879 & 1.890 $\pm$ 0.011 & 4.43 & [0.075 -- 19.842] \\
HS40 67\% 0$^\circ$ & 0.378 $\pm$ 0.014 & 43.617 $\pm$ 2.132 & 1.967 $\pm$ 0.012 & 3.15 & [0.094 -- 17.873] \\
HS40 62\% 0$^\circ$ & 0.309 $\pm$ 0.021 & 34.897 $\pm$ 3.203 & 1.960 $\pm$ 0.018 & 3.21 & [0.095 -- 18.531] \\
T300 67\% 0$^\circ$ & 0.198 $\pm$ 0.006 & 19.017 $\pm$ 0.967 & 2.082 $\pm$ 0.016 & 3.01 & [0.096 -- 19.799] \\
IMA 90$^\circ$ & 0.072 $\pm$ 0.007 & 19.108 $\pm$ 2.606 & 1.946 $\pm$ 0.023 & 5.03 & [0.134 -- 27.387] \\
IMA 0$^\circ$ & 0.533 $\pm$ 0.012 & 44.921 $\pm$ 1.698 & 2.166 $\pm$ 0.015 & 4.85 & [0.089 -- 22.620] \\
\hline
\end{tabular}
\end{sidewaystable*}
In equation~\ref{eq:fit_function_thermal_conductivity}, the parameter $a$ scales with the phonon relaxation time in the low-temperature scattering regime; larger values of $a$ correspond to higher intrinsic phonon conductance, consistent with higher fibre content or more graphitic fibre structure (as seen for M55J and IMA 0$^\circ$). The parameter $b$ governs the crossover between the $T^3$ Debye regime and the higher-temperature behavior. The exponent $n\approx 2$ for all samples is an empirical fitting parameter governing the crossover between the low-temperature $T^3$ Debye regime and the weaker power-law dependence observed at higher temperatures. Its physical interpretation in terms of a specific scattering mechanism requires further investigation and is beyond the scope of the present phenomenological fit.
\section{Comparison with literature}
\begin{figure*}[t]
	\includegraphics[width=\textwidth]{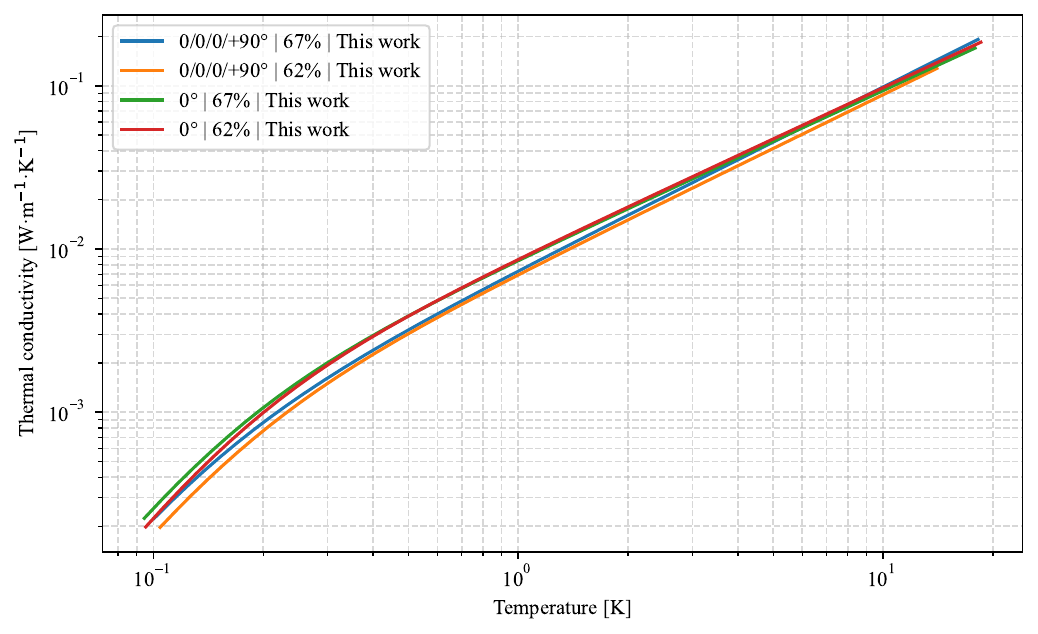}
	\caption{Thermal conductivities for the HS40 samples obtained in this work.}
	\label{fig:comparison_literature_HS40}
\end{figure*}

\begin{figure*}[t]
	\includegraphics[width=\textwidth]{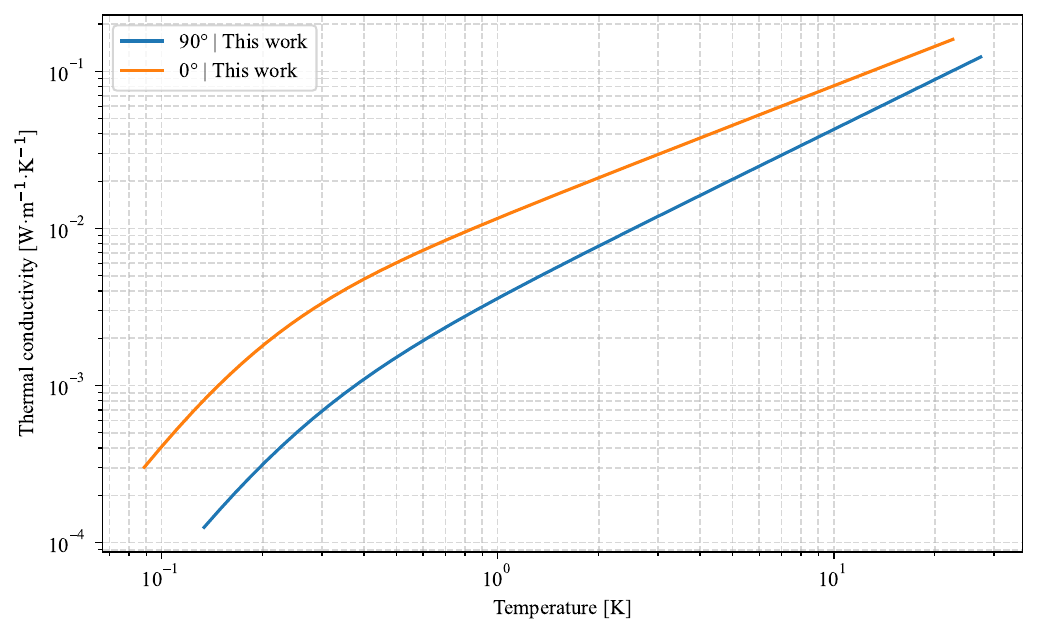}
	\caption{Thermal conductivities for the IMA samples obtained in this work.}
	\label{fig:comparison_literature_IMA}
\end{figure*}

Unfortunately, although a substantial number of thermal-conductivity measurements can be found in the literature, the corresponding results are often reported simply under generic labels such as ``CFRP'' or ``carbon-fibre composite'', without specifying the fibre type, the epoxy system, or the volume fraction.
In the following, we compare our measurements with the most appropriate data available in the literature. However, no published thermal-conductivity data exist for CFRPs based on HS40 or IMA fibres, which prevents any meaningful comparison for these specific materials. The thermal conductivities are presented in Figure~\ref{fig:comparison_literature_HS40} for the HS40 material and in Figure~\ref{fig:comparison_literature_IMA} for a more detailed comparison.
\subsection{T700}
\begin{figure*}[t]
	\includegraphics[width=\textwidth]{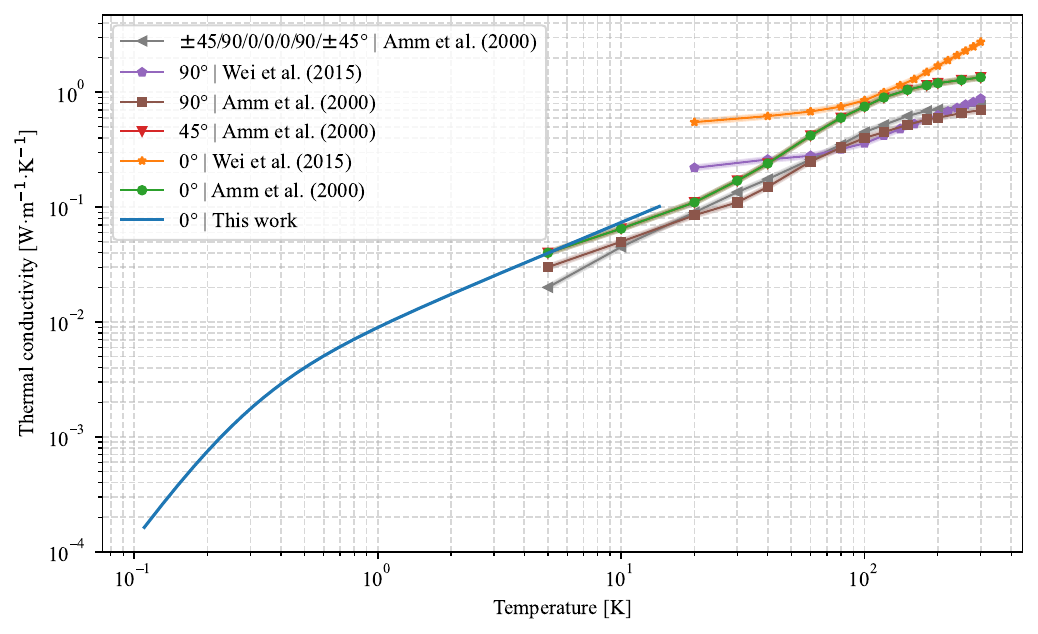}
	\caption{Comparison between the thermal conductivities for T700 obtained in this work and those reported in the literature over the entire temperature range, from room temperature down to the lowest temperatures.}
	\label{fig:comparison_literature_T700}
\end{figure*}
The present 0$^\circ$ measurements exhibit excellent agreement with the data of Amm~\textit{et~al.}~\cite{amm_thermal_2000} over the full overlapping temperature interval [5--15~K], both in absolute magnitude and in their temperature dependence, as shown in Figure~\ref{fig:comparison_literature_T700}. Conversely, the values reported by Wei~\textit{et~al.}~\cite{wei_cryogenic_2015} are systematically higher than our measurements: typically by a factor of 1.5--2 for the 0$^\circ$ orientation and by several tens of percent for the 90$^\circ$ direction. Such differences are not uncommon in the literature and are likely attributable to variations in material formulation, fibre volume fraction, lay-up, or experimental protocol. In the absence of a more complete description of the tested material, a definitive conclusion cannot be drawn.
\subsection{T300}
\begin{figure*}[t]
	\includegraphics[width=\textwidth]{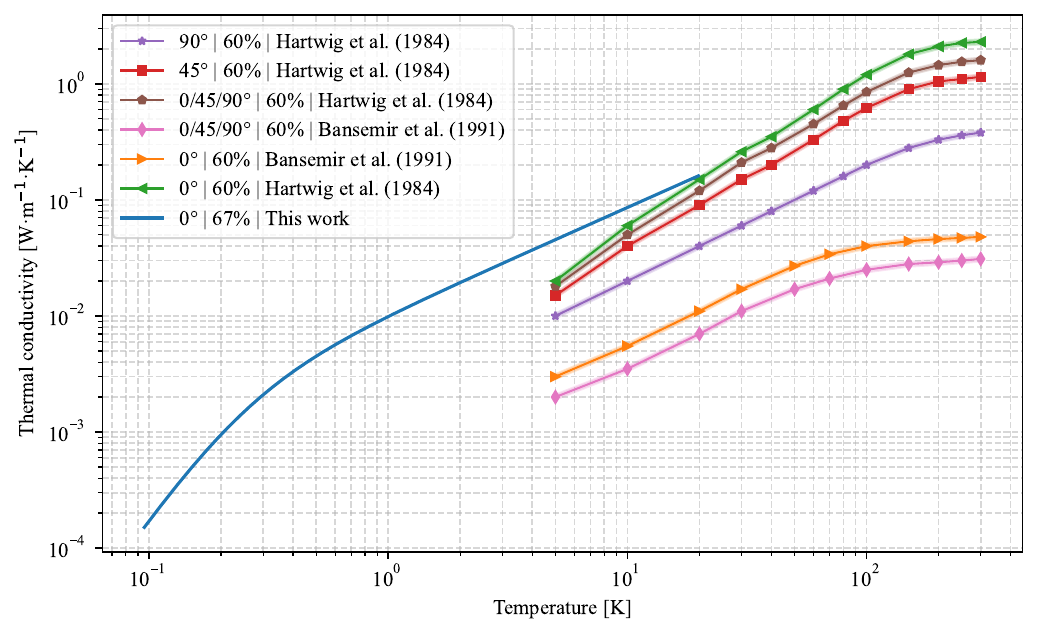}
	\caption{Comparison between the thermal conductivities for T300 obtained in this work and those reported in the literature over the entire temperature range, from room temperature down to the lowest temperatures.}
	\label{fig:comparison_literature_T300}
\end{figure*}
The present 0$^\circ$ measurements exhibit good agreement with the 0$^\circ$ data reported by Hartwig~\textit{et~al.}~\cite{hartwig_fibre-epoxy_1984}, as represented in Figure~\ref{fig:comparison_literature_T300}. The literature plots for 45$^\circ$, 90$^\circ$, and cross-plied laminates follow the expected hierarchy imposed by fibre orientation, with heat transport remaining most efficient along the fibre direction. In contrast, the values published by Bansemir~\textit{et~al.}~\cite{bansemir_basic_1991} are consistently lower than those of Hartwig~\textit{et~al.} for comparable orientations, which likely reflects differences in material formulation or experimental procedures. These discrepancies should be borne in mind when comparing the various datasets.
\subsection{M55J}
\begin{figure*}[t]
	\includegraphics[width=\textwidth]{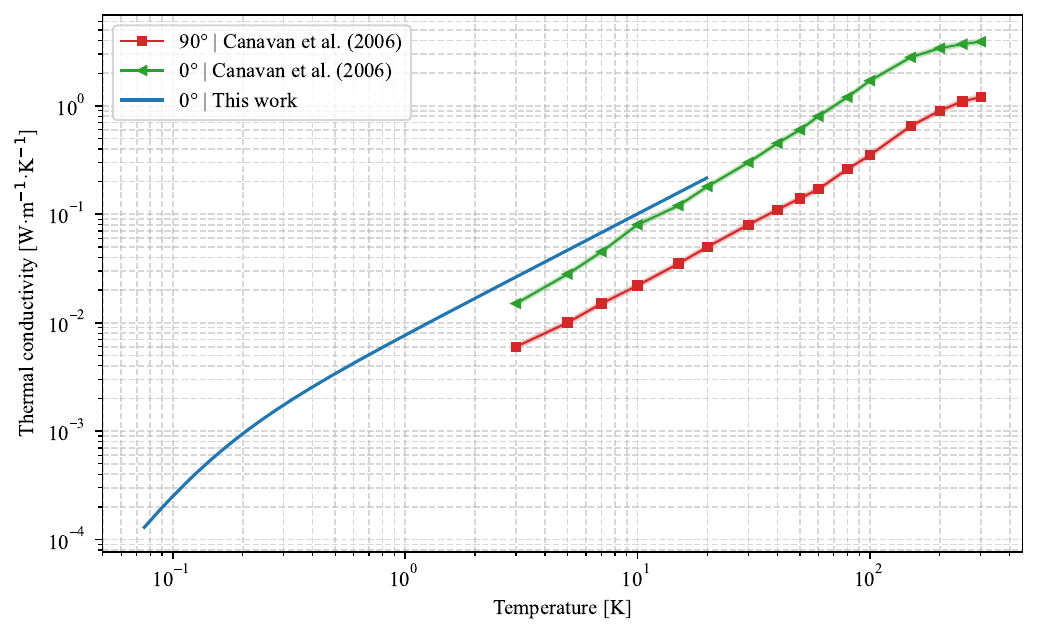}
	\caption{Comparison between the thermal conductivities for M55J obtained in this work and those reported in the literature over the entire temperature range, from room temperature down to the lowest temperatures.}
	\label{fig:comparison_literature_M55J}
\end{figure*}
The present 0$^\circ$ measurements reproduce the overall behaviour reported by Canavan~\textit{et~al.}~\cite{canavan_thermal_2006}, with comparable values above a few kelvin, as shown in Figure~\ref{fig:comparison_literature_M55J}. As expected, the fibre-parallel (0$^\circ$) data of Canavan~\textit{et~al.} remain systematically higher than their 90$^\circ$ measurements, illustrating the pronounced anisotropy of unidirectional CFRP. The slight offset observed between our 0$^\circ$ curve and the corresponding literature values is consistent with differences in fibre volume fraction, resin system, or laminate architecture, all of which exert a direct influence on the absolute level of thermal conductivity. Overall, both datasets exhibit a similar temperature dependence over the common interval, while the variations in magnitude can be readily attributed to differences in material composition and lay-up.
\section{Conclusion}
We measured the thermal conductivity $\kappa$ of nine CFRP samples with different fiber types, fiber volume fractions, and laminate structures, over temperatures from 100~mK to 20~K. The experimental setup and method developed for this work allow us to obtain reliable values down to the millikelvin range. The data show clear trends linked to fiber orientation and fiber content, and all samples follow the expected low-temperature phonon-scattering behavior.
Compared with the most relevant published results, our measurements agree well with datasets based on similar material definitions. The systematic differences seen in other studies emphasize how important it is to specify composite constituents and processing conditions accurately. The thermal-conductivity parametrization obtained here therefore offers a solid basis for modeling CFRP components in current and future cryogenic space instruments, especially at temperatures where no measurements previously existed.
Our results also show that, for all samples, an inflection in the conductivity curve below 1~K makes extrapolation from higher temperatures unreliable. The physical origin of this inflection likely involves a transition in the dominant heat-transport mechanism. Above $\sim$1~K, phonon conduction in the carbon fibres (governed by boundary and defect scattering with a $T^3$ dependence) dominates. Below $\sim$1~K, the phonon mean free path saturates at the fibre diameter or inter-defect spacing, and the relative contribution of the epoxy matrix, as well as fibre--matrix interface scattering,  becomes more significant. A detailed investigation of these mechanisms is left for future work. As a consequence, when CFRPs are used in critical systems operating at sub-Kelvin temperatures, each new material must be characterized. Thermal conductivity must be considered together with mechanical properties to select a suitable CFRP for the intended application.
\section*{Acknowledgements}
The author would like to thank the CEA--DSBT (Département des Systèmes Basses Températures, Grenoble, France), the IRAP (Institut de Recherche en Astrophysique et Planétologie, Toulouse, France), and MECANO~ID (Toulouse, France) for providing the samples. The author also wishes to thank Thomas Prouvé, Jean-Christophe Le~Clec'h and Antoine Arondel for their valuable assistance.
\bibliography{sn-bibliography}
\end{document}